\RequirePackage{etex}
\documentclass[VANCOUVER, Times1COL]{WileyNJD-v5}

\usepackage{graphicx}
\usepackage{amsmath}
\usepackage{amsfonts}
\usepackage{amssymb}
\usepackage[utf8]{inputenc}
\usepackage{url}
\usepackage{booktabs}
\usepackage{longtable}
\usepackage{parskip}
\usepackage{placeins}

\newcommand{\ith}[0]{\textsuperscript{th}}
\newcommand{\indep}{\perp \!\!\! \perp}
\DeclareMathOperator{\logit}{logit}

\PassOptionsToPackage{sort&compress}{natbib}

\PassOptionsToPackage{nopatch=footnote}{microtype}

\articletype{Research Article}
\received{00}
\revised{00}
\accepted{00}
\authormark{Gasparini et al.}
\titlemark{Analysis of stepped wedge trials with non-ignorable dropout}

\begin{document}

% Title
\title{Analysis of cohort stepped wedge cluster-randomized trials with non-ignorable dropout via joint modeling}

% Authors
\author[1,2]{Alessandro Gasparini}
\author[1]{Michael J. Crowther}
\author[3]{Emiel O. Hoogendijk}
\author[4,5]{Fan Li$^{\dagger,}$}
\author[6,7]{Michael O. Harhay$^{\dagger,}$}

\address[1]{Red Door Analytics AB, Stockholm, Sweden}
\address[2]{Department of Medical Epidemiology and Biostatistics, Karolinska Institutet, Stockholm, Sweden}
\address[3]{Department of Epidemiology and Data Science, Amsterdam Public Health Research Institute, Amsterdam UMC -- Location VU University Medical Center, Amsterdam, Netherlands}
\address[4]{Department of Biostatistics, Yale School of Public Health, Connecticut, USA}
\address[5]{Center for Methods in Implementation and Prevention Science, Yale School of Public Health, Connecticut, USA}
\address[6]{Department of Biostatistics, Epidemiology, and Informatics, Perelman School of Medicine at the University of Pennsylvania, Pennsylvania, USA}
\address[7]{MRC Clinical Trials Unit, University College London, London, UK}

\footnotetext{Correspondence to: Alessandro Gasparini, e-mail: \href{mailto:alessandro.gasparini@reddooranalytics.se}{alessandro.gasparini@reddooranalytics.se}. The $\dagger$ symbol denotes shared co-senior authorship.}

\abstract{Stepped wedge cluster-randomized trial (CRTs) designs randomize clusters of individuals to intervention sequences, ensuring that every cluster eventually transitions from a control period to receive the intervention under study by the end of the study period.
The analysis of stepped wedge CRTs is usually more complex than parallel-arm CRTs due to more complex intra-cluster correlation structures.
A further challenge in the analysis of closed-cohort stepped wedge CRTs, which follow groups of individuals enrolled in each period longitudinally, is the occurrence of dropout.
This is particularly problematic in studies of individuals at high risk for mortality, which causes non-ignorable missing outcomes.
If not appropriately addressed, missing outcomes from death will erode statistical power, at best, and bias treatment effect estimates, at worst.
Joint longitudinal-survival models can accommodate informative dropout and missingness patterns in longitudinal studies.
Specifically, within the joint longitudinal-survival modeling framework, one directly models the dropout process via a time-to-event submodel together with the longitudinal outcome of interest.
The two submodels are then linked using a variety of possible association structures.
This work extends linear mixed-effects models by jointly modeling the dropout process to accommodate informative missing outcome data in closed-cohort stepped wedge CRTs.
We focus on constant intervention and general time-on-treatment effect parametrizations for the longitudinal submodel and study the performance of the proposed methodology using Monte Carlo simulation under several data-generating scenarios.
We illustrate the joint modeling methodology in practice by reanalyzing data from the `Frail Older Adults: Care in Transition' (ACT) trial, a stepped wedge CRT of a multifaceted geriatric care model versus usual care in 35 primary care practices in the Netherlands.
}


\maketitle

\textcolor{red}{The published version of this manuscript can be accessed online at the following URL: \url{http://dx.doi.org/10.1002/sim.10347}}

\section{Introduction} \label{introduction}

\begin{figure}
\centering
\includegraphics[width = 0.7\textwidth]{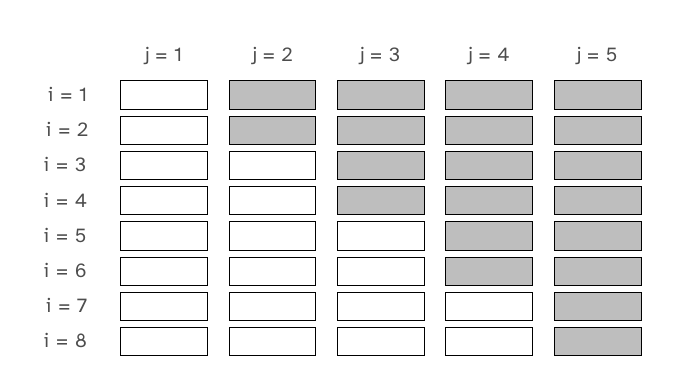}
\caption{Example of a closed-cohort stepped wedge trial with \(i = 8\) clusters and \(j = 5\) time periods; grey boxes represent being under study treatment during a given cluster period. Reproduced from Li et al. \cite{Li_2021} \label{fig:swt}}
\end{figure}

Cluster-randomized trials (CRTs) are popular study designs used to evaluate interventions delivered to a group of individuals, such as clinics, hospitals, or villages.
CRTs are adopted when the cluster is the intervention target, when individual randomization is infeasible, or when the risk for contamination is high. \cite{CRT}
In the most straightforward design, clusters in CRTs are randomized in parallel, with each cluster randomly assigned to either intervention or control with no crossover.
An alternative CRT design uses the so-called stepped wedge randomization, with clusters randomized to crossover from a control condition at different times, i.e., randomizing when the intervention condition will start. \cite{hussey_design_2007, Li_2021}
In this design, each cluster transitions to the intervention at a different point in time, and by the end of the trial, all clusters have received the intervention under study.
Benefits and tradeoffs of stepped wedge designs are discussed in detail elsewhere \cite{SWCRT, hemming_sample_2016, Li_2021}; among others, they overcome difficulties in rolling out the intervention to all study participants at the same time, they allow every study participant to be exposed to the study intervention(s), and they can, in some settings, increase statistical power as each cluster serves as its own control.
For instance, Hemming and Taljaard compared the relative efficiency of parallel CRTs and stepped wedge CRTs and concluded that statistical power depends on within-cluster correlation and cluster size, with stepped wedge CRTs being more efficient than parallel CRTs when the intra-cluster correlation is high. \cite{hemming_sample_2016}
Moreover, there are several types of stepped wedge designs, including open, cross-sectional, and closed-cohort, each referring to how participants are followed in the study. \cite{copas_designing_2015, nevins2023adherence}
The work focuses on inferential methods for non-ignorable missing outcome data due to death in closed-cohort designs, where the outcomes of individuals are measured repeatedly over the study period (Figure \ref{fig:swt}).

The loss of study participants in longitudinal studies is a major challenge that, if not adequately handled, can introduce bias in the analysis and affect the identification of true treatment effects.
This is especially relevant when the loss to follow-up is correlated to socioeconomic position or poor health, which can correlate with a higher risk of dropping out. \cite{young_attrition_2006}
For instance, a recent review of nursing home CRTs by Poupin et al. found that a median of 19.5\% study participants did not complete the entire study protocol, mostly because of death. \cite{poupin_management_2022}

Dropout in longitudinal studies is often ignored or inadequately examined to assess the potential impact of missing outcome data (and approaches to handling missing data) on study conclusions. \cite{poupin_management_2022}
In this manuscript, we approach this problem by explicitly modeling the dropout process together with the longitudinal outcome within the shared random-effects modeling framework.
Other approaches, such as locating and re-engaging study participants who were lost to follow-up, have been proposed in the literature \cite{bennetts_selective_2021}; this is, however, not always possible, particularly when dropout is due to the death of the study participant.
Our approach is tailored to the settings of closed-cohort stepped wedge CRTs but could be extended to parallel and cluster-crossover CRTs (and other multilevel settings), when repeated measurements are collected over multiple time periods, as well.

\begin{figure}
\centering
\includegraphics[width = 0.7\textwidth]{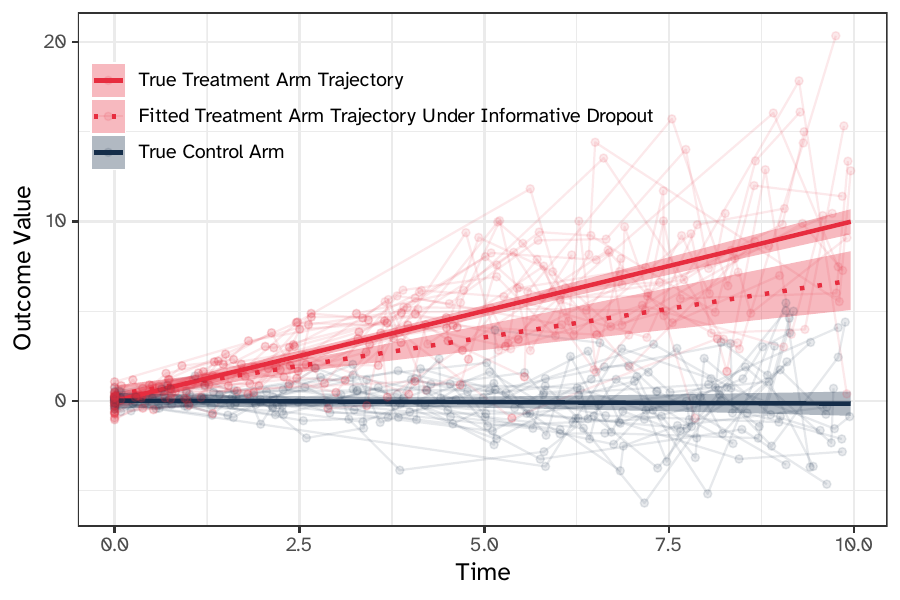}
\caption{Illustrative example of biased longitudinal trajectories in the settings of informative dropout, using simulated data. The example assumes that study participants drop out when the longitudinal outcome exceeds the value of five, e.g., representing a hypothetical scenario where participants have improved enough to no longer perceive the benefits of continuing the study. \label{fig:swt_infdropout}}
\end{figure}

It is known in the literature that dropout from the study, per se, does not necessarily introduce bias in the analysis.
For instance, if dropout corresponds to outcome missing completely at random (MCAR), then the dropout process is ignorable and only leads to loss of efficiency.
In studies where MCAR is expected, inflation of the planned sample size is a convenient strategy to restore study power, and standard methods for the analysis of longitudinal data can be used without bias. \cite{little1987statistical, poupin_management_2022, rouanet_interpretation_2019, rouanet_how_2022}
Mixed-effects models estimated on the available data are recommended in these settings, as their maximum likelihood estimators are robust to missing at random (MAR); conversely, generalized estimating equations (GEE) estimators are only unbiased when data is MCAR, but a weighted GEE can be used to overcome this limitation under MAR. \cite{rouanet_interpretation_2019, turner2020properties, rouanet_how_2022}
When dropout is not MCAR nor MAR, then it is said to be missing not at random (MNAR), informative, or non-ignorable: in these settings, the selection bias over time induced by differential attrition rates (e.g., between treatment arms) can lead to bias.
We illustrate this with a simulated example in Figure \ref{fig:swt_infdropout}, where we randomized 50 patients to a binary treatment (or control) and assume a non-null treatment effect (i.e., treated subjects improve/increase their longitudinal outcome over time).
We also assume that study subjects leave the study once their outcome exceeds the value of five, e.g., because they have improved enough to no longer perceive the benefits of continuing the study.
In this setting, we would underestimate the benefit of the treatment if we were ignoring this informative dropout (red dotted line) compared to the true underlying longitudinal trajectory (red solid line), with an estimated linear slope in the observed treatment arm of 0.64 (95\% C.I.: 0.44, 0.84) instead of a true slope of 1.
Note that, in the MNAR settings, attrition rates may depend on unobserved characteristics of the longitudinal outcome, such as the current true value or the slope of change, and that dropout from the study may be informative even with equal attrition rates between treatment arms, i.e., if dropout occurs for different reasons, such as toxicity in the treatment arm and lack of perceived benefit in the control arm.
This leads to a form of informative censoring that is cryptic and hard to identify. \cite{olivier_equal_2024}
The difference between MCAR, MAR, and MNAR settings in randomized trials is further discussed, among other methodological considerations, by Thakur et al. \cite{thakur_statistical_2023}

The rest of this manuscript is organized as follows.
In Section \ref{review}, we review and discuss the current literature on informative dropout in trials with a longitudinal outcome.
In Section \ref{mixed}, we introduce the mixed modeling parametrization for closed-cohort stepped wedge CRTs that we focus on throughout this work.
In Section \ref{joint-model}, we extend the mixed modeling approach to the joint longitudinal-survival framework, including model formulation, likelihood, and estimation.
In Section \ref{mcsim}, we compare the two approaches using Monte Carlo simulation methods.
In Section \ref{act}, we re-analyze data from the Frail Older Adults: Care in Transition (ACT) trial, a pragmatic stepped wedge CRT of a multifaceted geriatric care model versus usual care in 35 primary care practices in the Netherlands, using both the traditional mixed modeling and the joint modeling approach.
In Section \ref{extensions}, we discuss extensions to the proposed framework.
We conclude the manuscript with a discussion in Section \ref{discussion}.

\section{Literature review and objectives} \label{review}

The issue of informative dropout in trials with longitudinal outcomes, and the role of joint modeling specifically, has been previously studied in the literature.
Rouanet et al. summarized different approaches for modeling longitudinal outcomes and highlighted how joint longitudinal-survival models can accommodate MNAR settings, while mixed models and (potentially weighted) GEE models cannot. \cite{rouanet_how_2022}
Kolamunnage-Dona et al. illustrated the joint modeling approach in randomized trials (without any clustering) with competing causes of dropout, highlighting the value of this approach for evaluating the sensitivity of conclusions to assumptions regarding missing data mechanisms. \cite{kolamunnage-dona_modelling_2016}
Cuer et al. compared linear mixed models and joint longitudinal-survival models to show their differences in terms of outputs, interpretation, and underlying modeling assumptions. \cite{cuer_handling_2020}
In further work, they compared standard joint longitudinal-survival models with competing risks joint models (e.g., those discussed by Kolamunnage-Dona et al. as well \cite{kolamunnage-dona_modelling_2016}) both in practice and via simulation, showing that the standard model and the competing risks model performed equally well when the risk of dropout was the same whatever the cause. \cite{cuer_joint_2021}
Moreover, they showed that the standard joint model led to biased results when the risk of dropout differed between causes; this was not the case for the competing risks joint model. \cite{cuer_joint_2021}
Hogan et al. developed a mixture-modeling approach for longitudinal data with outcome-dependent dropout where the functional dependence between covariate effects and dropout time can be left unspecified \cite{Hogan_2004}; their approach can be considered to be an extension of selection models. \cite{diggle_informative_1994, wu_estimation_1988}
Wang and Chinchilli developed a shared parameters model for the settings of crossover designs with informative dropout. \cite{Wang_2020}
Specifically, they combine mixed-effects models with a model for dropout based on fully parametric discrete hazard models, and they implement maximum likelihood estimation to estimate the model parameters.
Moreover, they extended their approach to binary longitudinal outcomes and showed that GEE models and generalized linear mixed models with pseudo-likelihood estimation for complete and incomplete data were robust to informative dropout, highlighting that, in their simulation settings, the impact from the missing data on the model estimates was mild with a binary outcome. \cite{wang_analysis_2022}

We now describe and compare in more detail the three main approaches that have been proposed for modeling MNAR outcomes: pattern mixture models, selection models, and joint longitudinal-survival models.
First, pattern mixture models factorize the joint distribution of the longitudinal outcome and missingness mechanism into the probability of missing data and the conditional distribution of the outcome given missingness.
One such example for analyzing continuous outcomes in longitudinal cluster randomized trials was developed by Fiero et al., where they combined pattern mixture models with multilevel multiple imputation to create MNAR imputed values. \cite{fiero_patternmixture_2017}
This approach provides a viable strategy for performing sensitivity analyses under certain user-defined assumptions regarding the missing data mechanism.
Demirtas and Schafer discuss random coefficient pattern mixture models, that is, random-effects models for a certain longitudinal outcome that include summaries of the dropout mechanism as fixed effects. \cite{demirtas_performance_2003}
In their work, they illustrate how alternative random coefficient pattern mixture models that fit the data equally well may lead to very different estimates for the parameters of interest, highlighting how even minor model misspecification can introduce large biases.
Second, selection models factorize the joint distribution of longitudinal outcome and missingness the other way around, as a model for the study outcome (the substantive analysis) times the conditional probability of missingness given the outcome; specifically, the Diggle-Kenward selection model combines a growth curve model (as the substantive model) with a set of logistic or probit regression models (as the conditional distribution) that directly relate the missing data probability to the outcome variable at current and/or previous time points. \cite{diggle_informative_1994}
An extension of selection models is given by Hogan et al. \cite{Hogan_2004}
Moreover, some of these approaches have been previously compared via Monte Carlo simulation, concluding that selection models outperformed pattern mixture models --- especially as sample size and dropout rate increased. \cite{chen_comparison_2020}
Third, the shared random effects, joint modeling approach introduces a common set of latent random effects that link the longitudinal outcome and the dropout mechanism. \cite{wulfsohn_joint_1997, ibrahim_basic_2010, gould_joint_2015}
The longitudinal and survival outcomes are then assumed to be conditionally independent (given the random effects) and are jointly estimated within a unified framework, with flexibility in terms of model formulations for the two submodels.
An example is given by Wang and Chinchilli in the context of crossover trials. \cite{Wang_2020}

The three approaches for MNAR outcomes described above have advantages and disadvantages.
Pattern mixture models and selection models can have relatively simple formulations and are usually easier to estimate compared to joint longitudinal-survival models.
However, these approaches require distributional assumptions for model parameters to be identifiable, and it has been shown that fitted parameters may vary widely under different missing data assumptions. \cite{demirtas_performance_2003, fiero_patternmixture_2017, chen_comparison_2020}
On the other end, shared random-effects joint models are computationally more challenging to estimate, and the population-level interpretation of the fitted coefficients for the outcome model is only valid in an immortal cohort. \cite{rouanet_interpretation_2019}
Nonetheless, we choose to focus on this approach for a variety of reasons.
First, joint longitudinal-survival models provide much more flexibility (in terms of possible model formulations) currently available in statistical software packages, with further improvements as software implementations develop.
Second, this class of models is much closer to standard approaches for the analysis of stepped wedge CRTs based on mixed-effects models, \cite{Li_2021} as they extend the usual methods by explicitly modeling the dropout process on top of the outcome model.
Therefore, we believe these methods to be much more approachable by practitioners.
Third, joint longitudinal-survival models do not significantly contaminate the conventional model-based target treatment effect parameters in stepped wedge CRTs, whereas other methods may do so without careful reparameterization.

Finally, ther design approaches to tackle the issue of informative dropout are also possible. For instance, Poupin et al. suggested that open-cohort designs could be used to reduce the impact of potentially informative attrition, as long as (1) the intervention and the main cause of attrition are independent and (2) the intervention has a cluster-specific effect (if any). \cite{poupin_cluster_2023}
This is, therefore, not feasible with interventions that are expected to have an individual effect, such as medications or treatments administered at the subject level, and is beyond the scope of this manuscript.
Despite this existing work, there has been limited research on methods for addressing informative attrition in closed-cohort stepped wedge CRTs, where the exogenous, time-dependent treatment indicator can affect both the attrition rate and the longitudinal outcome.
In addition, recent studies have indicated the importance of accounting for time-dependent treatment effects in the longitudinal modeling in stepped wedge CRTs, and it remains of interest to develop methods in the presence of both time-dependent treatment effects and informative attrition. \cite{kenny_analysis_2022, maleyeff_assessing_2023, wang2024achievemodelrobustinferencestepped}

%%% ---------------

\section{Mixed-effects models: a recap} \label{mixed}

Closed-cohort stepped wedge trials are routinely analyzed using mixed-effects models. \cite{Li_2021}
For this project, we focus on the following linear mixed-effects model:
\begin{equation}
  \label{eq:sm-ne-ci}
  \begin{aligned}
  Y_{ijk}(t) &= m_{ijk}(t) + \varepsilon_{ijk}(t) = \sum_{j=1}^J \beta_j I(T_{j - 1} < t \le T_j) + \delta X_i(t) + \alpha_i + \phi_{ik} + \varepsilon_{ijk}(t)
  \end{aligned}
\end{equation}
where \(Y_{ijk}(t)\) is a continuous outcome measured at time \(t\) for the i\ith{} cluster during the j\ith{} discrete period and on the k\ith{} study participant.
\(X_i(t)\) is a binary covariate denoting whether the i\ith{} cluster received treatment during the discrete calendar period \((T_{j - 1}, T_j]\), \(\delta\) is the treatment effect, and \(\beta_j\) are period-effect coefficients.
We assume that \(\varepsilon \sim N(0, \sigma^2_{\varepsilon})\) is residual (measurement) error, and we include two independent random effects, \(\alpha \sim N(0, \sigma^2_{\alpha})\) and \(\phi \sim N(0, \sigma^2_{\phi})\), representing a cluster-, and a subject-specific random intercept, respectively.
This model is sometimes referred to as a \emph{nested exchangeable} random-effects model, with an immediate and constant intervention effect; it represents an extension of the Hussey and Hughes model introduced by Baio et al. \cite{Baio_2015}
A nested exchangeable model implies the following correlation structure:
$$
\text{corr}[Y_{ijk}(t), Y_{ilm}(t)] =
\begin{cases}
  \rho_a = \displaystyle \frac{\sigma^2_{\alpha} + \sigma^2_{\phi}}{\sigma^2_{\alpha} + \sigma^2_{\phi} + \sigma^2_{\varepsilon}}, & k = m \\
  \rho_d = \displaystyle \frac{\sigma^2_{\alpha}}{\sigma^2_{\alpha} + \sigma^2_{\phi} + \sigma^2_{\varepsilon}}, & k \ne m
\end{cases}
$$
where $\rho_a$ denotes the within-individual intra-class correlation coefficient (ICC) and $\rho_d$ the correlation between two observations collected from different individuals, irrespective of time periods.

\begin{figure}
  \centering
  \includegraphics[width = 0.7\textwidth]{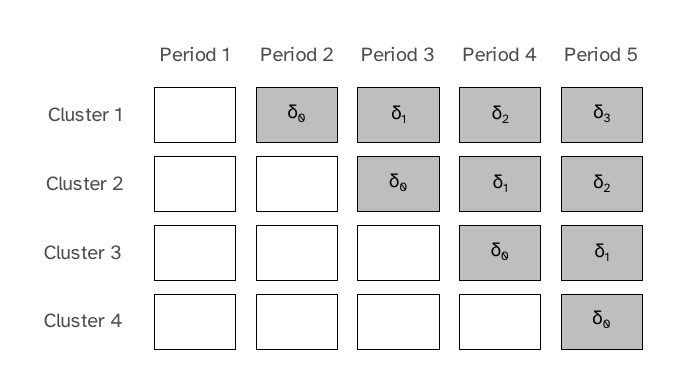}
  \caption{An example of a closed-cohort stepped wedge trial with \(i = 4\) clusters and \(j = 5\) time periods, with a general time on treatment effect \(\delta_{j-s}\); grey boxes represent intervention periods, while white boxes represent control periods. \label{fig:swt-gtot}}
\end{figure}

We can generalize the constant intervention model of Equation \eqref{eq:sm-ne-ci} by allowing for a general time on treatment effect:
\begin{equation}
  \label{eq:sm-ne-gi}
  \begin{aligned}
  Y_{ijk}(t) &= m_{ijk}(t) + \varepsilon_{ijk}(t) = \sum_{j=1}^J \beta_j I(T_{j - 1} < t \le T_j) + \delta_{j-s} X_i(t) + \alpha_i + \phi_{ik} + \varepsilon_{ijk}(t)
  \end{aligned}
\end{equation}
where $\delta_{j-s}$ denotes an intervention effect that depends on how many time periods have passed since the intervention was introduced,
$$
\delta_{j-s} = \delta_0 I(j = s) + \delta_1 I(j = s + 1) + \delta_2 I(j = s + 2) + \cdots,
$$
assuming the intervention was introduced during period $s$.
This is illustrated in Figure \ref{fig:swt-gtot}.
This model estimates a potentially distinct treatment effect per level of the exposure time (duration of the treatment and is different from the calendar time), and addresses exposure-time-dependent treatment effect heterogeneity, which may arise due to the learning, weakening, or delayed effect due to implementation of the intervention. \cite{Li_2021, kenny_analysis_2022, maleyeff_assessing_2023, wang2024achievemodelrobustinferencestepped}

%%% ---------------

\section{A joint modeling approach to address informative attrition in cohort stepped wedge designs} \label{joint-model}

When dropout from the study is informative, the traditional mixed-effects analysis models introduced in Section \ref{mixed} can yield biased parameter estimates as it assumes outcomes are MAR, i.e., that missingness is non-informative.
To overcome this limitation, we propose to simultaneously model the longitudinal and dropout processes via shared random-effects longitudinal-survival modeling.

We now introduce additional notation to describe the dropout process.
Specifically, let \(S_{ik}\) be the full observation time for the k\ith{} participant in the i\ith{} cluster.
This corresponds to the full time under study, e.g., until the end of the final intervention period, for a patient that completes the study.
However, study participants can drop out before the end of the last intervention period for various reasons, including death or migration from the country.
Therefore, in practice, we can only observe time \(T_{ik} = \min(S_{ik}, C_{ik})\), where \(C_{ik}\) is the dropout time; the indicator variable \(\Delta_{ik} = I(C_{ik} \le S_{ik})\) represents whether a participant dropped out of the study before the end of follow up (\(\Delta_{ik} = 1\)) or not (\(\Delta_{ik} = 0\)). The observed time to event data will thus be, for the k\ith{} participant in the i\ith{} cluster, the couple \((T_{ik}, \Delta_{ik})\).

\subsection{Joint model formulation} \label{joint-model:formulation}

We propose the following general joint model, which extends the constant intervention effect, nested exchangeable model:
\begin{equation}
\label{eq:jm-ci}
\begin{cases}
Y_{ijk}(t) = \sum_{j=1}^J \beta_j I(T_{j - 1} < t \le T_j) + \delta X_i(t) + \alpha_i + \phi_{ik} + \varepsilon_{ijk}(t) \\
\lambda_{ijk}(t) = \lambda_0(t) \exp(\nu X_i(t) + \omega Z_{ik})
\end{cases}
\end{equation}
where $Z_{ik} = [\alpha_i, \phi_{ik}]$ denotes any user-specified combination of the random effects $\alpha$, $\phi$ with associated regression coefficients $\omega$; the dimension of $\omega$ depends on which random effects from the longitudinal submodel are included in $Z_{ik}$.
Note that the association parameter(s) $\omega$ are sometimes referred to as \emph{factor loadings}. \cite{bollen1989structural}
We denote this parametrization of the association structure between the longitudinal and time-to-dropout models as the \emph{shared random-effects parametrization}.
Moreover, the survival submodel assumes a proportional hazards model with baseline hazard function \(\lambda_0(t)\) and treatment effect \(\nu\), but other covariates could be incorporated in the dropout model, in principle.

If we assume a maximal model from Equation \eqref{eq:jm-ci}, we obtain the following joint model:
\begin{equation}
  \label{eq:jm-ci-max}
  \begin{cases}
    Y_{ijk}(t) = \sum_{j=1}^J \beta_j I(T_{j - 1} < t \le T_j) + \delta X_i(t) + \alpha_i + \phi_{ik} + \varepsilon_{ijk}(t) \\
    \lambda_{ijk}(t) = \lambda_0(t) \exp(\nu X_i(t) + \omega_1 \alpha_i + \omega_2 \phi_{ik})
  \end{cases}
\end{equation}
where we allow for cluster- and subject-specific associations with the rate of dropout from the study (with association parameters \(\omega_1, \omega_2\)). We can simplify this maximal model by including constraints: for instance, we could constrain the effect of cluster- and subject-specific random effects on dropout to be the same (i.e., \(\omega_1 = \omega_2\)) or that a certain effect is zero (e.g., \(\omega_2 = 0\)).

Moreover, as in Section \ref{mixed}, we can extend the model from Equation \eqref{eq:jm-ci} by assuming a general time on treatment model for the longitudinal submodel:
\begin{equation}
  \label{eq:jm-gi}
  \begin{cases}
  Y_{ijk}(t) = \sum_{j=1}^J \beta_j I(T_{j - 1} < t \le T_j) + \delta_{j - s} X_i(t) + \alpha_i + \phi_{ik} + \varepsilon_{ijk}(t) \\
  \lambda_{ijk}(t) = \lambda_0(t) \exp(\nu X_i(t) + \omega Z_{ik})
  \end{cases}
\end{equation}

Note that we still assume a constant intervention effect on dropout, $\nu$, but this assumption could be relaxed as well, in principle.

Moreover, note that we assume continuous time for the dropout component of the joint model, and assume a parametric shape for the baseline hazard function $\lambda_0(t)$, such as exponential or Weibull.
For instance, an exponential baseline hazard function has a constant hazard defined by a single parameter $\lambda$, with $\lambda_0^{\text{Exp}}(t) = \lambda$, while a Weibull baseline hazard function with scale and shape parameters $\lambda$ and $p$, respectively, is defined as $\lambda_0^{\text{Wei}}(t) = \lambda p t^{p - 1}$.
Other parametric (e.g., Gompertz) or flexible parametric (e.g., using smooth splines to model the baseline hazard) formulations can be accommodated as well within the framework, in principle. \cite{royston_flexible_2002, Rubio_2019}
Semi-parametric and fully non-parametric baseline hazard functions are also possible, in principle, but that can lead to underestimation of the standard errors of parameter estimates requiring the bootstrap to obtain appropriate standard errors. \cite{hsieh_joint_2006}
The continuous-time modeling strategy also differentiates this work from that of Wang and Chinchilli, who considered a survival submodel based on a discrete-time formulation. \cite{Wang_2020}

Furthermore, we assume that dropout times are known exactly: in this scenario, we only have to consider right censoring for the time to dropout distribution.
In some settings, information that people have dropped out of the study might be only known at the end of each study period; in that scenario, the distribution for time to dropout would be interval censored, and the joint modeling approach would have to be extended to take this into account. \cite{zhigang_zhang_interval_2010}

Finally, simulating data from the general joint model formulation of Equations \eqref{eq:jm-ci} and \eqref{eq:jm-gi} is a non-trivial task given that it combines a longitudinal outcome measured with error and a time-varying, exogenous covariate (the randomized treatment sequence).
The algorithm for simulating data for our subsequent numerical study is described in more detail in the supplementary material.

\subsection{Likelihood and estimation} \label{joint-model:likelihood}

The joint model introduced in Section \ref{joint-model:formulation} can be fit using readily available statistical software within the maximum likelihood framework.
Specifically, the likelihood function is defined as follows:
\begin{equation}
  \label{eq:likelihood}
  \mathcal{L}(\theta) = \int_{\alpha} \int_{\phi} f(\mathbf{y} | \mathbf{x}, \alpha, \phi, \theta) \ f(a | \sigma^2_{\alpha}) \ f(p | \sigma^2_{\phi}) \ da \ dp
\end{equation}
where $\theta$ denotes the vector of model parameters, $\mathbf{y}$ the vector of response variables (the longitudinal and time to dropout outcomes), $\mathbf{x}$ the vector of model covariates (such as the fixed treatment and period effects), $f(\cdot | \sigma^2_{\alpha})$ the distribution of the cluster-specific random intercept $\alpha$, and $f(\cdot | \sigma^2_{\phi})$ the distribution of the subject-specific random intercept $\phi$.
The joint distribution of the random effects can be factorized as the product of their univariate distributions thanks to the independence assumption ($\alpha \indep \phi$) that follows from the nested exchangeable model formulation.
The integration in Equation \eqref{eq:likelihood} does not have closed-form, and needs to be approximated; to this end, we use mean-variance adaptive Gauss-Hermite quadrature as implemented in Stata's \texttt{gsem} command. \cite{pinheiro_approximations_1995, Stata}

A standard assumption in this class of models is that, conditionally on the random effects, the longitudinal measurements are independent of dropout times, i.e., $Y_{ijk} \indep T_{ik} \ | \ \alpha, \phi$ (using the notation of Section \ref{joint-model:formulation}).
Therefore, we can factorize the joint distribution of the longitudinal measurements and dropout times as well:
\begin{equation}
  f(\mathbf{y} | \mathbf{x}, \alpha, \phi, \theta) = f(\mathbf{Y} | \mathbf{x}^Y, \alpha, \phi, \theta^Y) \times f(\mathbf{t} | \mathbf{x}^t, \alpha, \phi, \theta^t)
\end{equation}
where $\mathbf{Y}$ denotes the longitudinal outcome and $\mathbf{t}$ the time to dropout outcome.
The vector of fixed effects covariates is partitioned into covariates for the longitudinal submodel ($\mathbf{x}^Y$) and for the survival submodel ($\mathbf{x}^t$), and analogously the vector of model parameters $\theta = (\theta^Y, \theta^t)$.
In practice, for our two-level model, the likelihood is computed at the cluster level by multiplying the subject-specific contribution of each subject in each cluster first and then combining all cluster contributions.

The likelihood function does not have a closed-form solution and is thus maximized numerically, e.g., using Stata's implementation of the Broyden-Fletcher-Goldfarb-Shanno (BFGS) algorithm or its modified Newton-Raphson algorithm. \cite{mlbook}
We implement the entire model in Stata, including Gaussian quadrature and optimization, within the generalized structural equation modeling framework using the \texttt{gsem} command. \cite{Stata}
More details on the estimation of hierarchical joint models can be found in Brilleman et al. \cite{brilleman_joint_2019}

%%% ---------------

\section{Monte Carlo simulation studies} \label{mcsim}

The setup of our simulation studies is described in this Section, following the ADEMP framework introduced by Morris et al. \cite{Morris_2019}

\subsection{Aims} \label{mcsim-aims}

Our simulation studies sought to assess the performance of traditional analysis methods for stepped wedge trials (introduced in Section \ref{mixed}) compared to the joint modeling approach outlined in Section \ref{joint-model}, under a variety of realistic scenarios.
Specifically, we design and perform two studies:

\begin{enumerate}
  \item Simulation study \emph{A}, where we study the performance of the constant intervention model described in Equation \eqref{eq:sm-ne-ci};
  \item Simulation study \emph{B}, where we study the performance of the general time on treatment model described in Equation \eqref{eq:sm-ne-gi}.
\end{enumerate}

\subsection{Data-generating mechanisms} \label{mcsim-dgm}

\begin{figure}
  \centering
  \includegraphics[width = 0.7\textwidth]{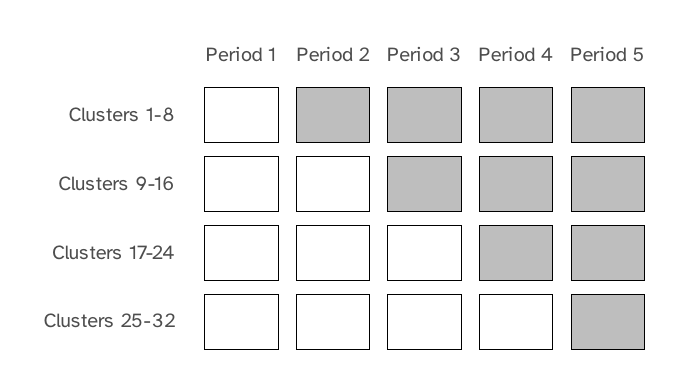}
  \caption{Closed-cohort stepped wedge trial design with 32 clusters (8 per intervention sequence, 4 intervention sequences) and 5 time periods, used for the simulation studies described in this manuscript; grey boxes represent being under study treatment during a given cluster period. \label{fig:swt-sim}}
\end{figure}

Characteristics of the data-generating mechanisms largely overlap across simulation studies \emph{A} and \emph{B}.
Specifically, we simulate a closed-cohort stepped wedge trial with 4 intervention sequences, 5 time periods, and 8 clusters per intervention sequence, for a total of 32 clusters, which mimics our application in Section \ref{act}; this is illustrated in Figure \ref{fig:swt-sim}.

Then, we assume that each cluster recruits 50 subjects at the beginning of the study, and that these subjects either complete the study or drop out before all intervention periods can be completed.
The dropout mechanism is consistent with the joint model described in Section \ref{joint-model} (and including the standard models of Section \ref{mixed} as a special case), and is described in more detail below.
Note that we assume that dropout is an irreversible absorbing state: i.e., once subjects drop out of the study, they cannot return under observation.

We assume that the time to dropout submodel follows a Weibull proportional hazards model with shape and scale parameters $p = 1.0$ and $\lambda = \exp(-1.5)$.
These values are based on the real-data application that is described in Section \ref{act}.
Then, we assume that the five parameters for the longitudinal period effects are $\beta_1 = \beta_2 = \beta_3 = \beta_4 = \beta_5 = 30$.
For the variance components, we assume a residual error with variance $\sigma^2_{\varepsilon} = 40$, and random intercepts variances of $\sigma^2_{\alpha} = 2$ and $\sigma^2_{\phi} = 55$.
These values correspond to a within-individual ICC ($\rho_a$) of 0.588 and a between-individuals ICC ($\rho_d$) of 0.021 which are within the range of commonly reported correlation parameters in stepped wedge designs. \cite{korevaar_intra_cluster_2021}

Under our simulation scenarios, we define a fully factorial design for the following model parameters:
\begin{itemize}
  \item The treatment effect(s) on the longitudinal outcome $\delta$, with $\delta \in \{0.0, 5.0, 25.0\}$ for the constant intervention model (simulation study \emph{A}) or $\{(\delta_0 = \delta_1 = \delta_2 = \delta_3 = 0.0), (\delta_0 = 0.00, \delta_1 = 2.50, \delta_2 = 5.00, \delta_3 = 6.25), (\delta_0 = 0.00, \delta_1 = 12.50, \delta_2 = 25.00, \delta_3 = 31.25)\}$ for the general time on treatment model (simulation study \emph{B}).
  Note that $5.0$ and $25.0$ represent approximately 5\% and 25\% of the total variance of the longitudinal outcome;
  \item The treatment effect on dropout $\nu$, with possible values $\nu \in \{-0.2, 0.0\}$.
  These coefficients correspond to hazard ratios for treated versus non-treated subjects of $0.819$ and $1.000$, respectively;
  \item The association parameters $\omega_1 = \omega_2 \in \{\log(0.5), \log(0.9), \log(1.0), \log(2.0)\}$.
  These correspond to hazard ratios of $0.5$, $0.9$, $1.0$, and $2.0$, respectively, and we deem scenarios with $\omega_1 = \omega_2 \in \{\log(0.5), \log(2.0)\}$ as extremes to demonstrate the performance of different methods in challenging scenarios.
\end{itemize}

Overall, the data-generating mechanisms here described consist of $3 \times 2 \times 4 = 24$ distinct scenarios, of which 6 (i.e., those with $\omega_1 = \omega_2 = \log(1.0) = 0.0$) correspond to settings where dropout from the study is non-informative.

Note that simulating time to dropout in continuous time, under this data-generating mechanism and with a time-varying exogenous exposure included in the survival submodel (the time-varying, deterministic treatment assignment), is a non-trivial task.
The algorithm we implemented is described in more detail below.

The procedure to simulate data from the joint model was developed by combining previous work by Bender et al., Austin, and Crowther and Lambert, and consists of two steps. \cite{Bender_2005, Austin_2012, Crowther_2013}
The first step consists of simulating the longitudinal data from a stepped wedge trial in the absence of dropout; then, we simulate the time to dropout from the study and censor the longitudinal trajectories accordingly.
These steps are formalized in Algorithm \ref{alg:simjm}.
Simulating dropout times from the survival submodel of Equations \eqref{eq:jm-ci}, \eqref{eq:jm-gi} follows from Bender et al., and is based on the inversion method.  \cite{Bender_2005}
Specifically, let \(T^s\) be the simulated dropout time; assuming that
$$
F(T^s | X, t, \alpha, \phi, \theta) = 1 - \exp[-H(T^s | X, t, \alpha, \phi, \theta)] = u,
$$
with $u \sim \text{Unif}(0, 1)$, we can solve for $T^s$ and obtain simulated dropout times.
However, in our case, this is not straightforward as the cumulative hazard function depends on (1) the time-varying treatment, (2) the time-period-specific effects, and (3) the random effects.
We can thus implement the procedure described by Austin: \cite{Austin_2012} assuming that $t_J$ is the start of the time interval where a given subject starts to be exposed to treatment, a Weibull baseline hazard with shape and scale parameters $\lambda$ and $p$, respectively, and $b = \lambda \exp(\omega_1 \alpha_i + \omega_2 \phi_{ik}) {t_J}^{p}$, dropout times can be simulated as
\begin{equation}
  \label{eq:sim-weibull-simple}
  T_i = \begin{cases}
    \left( \displaystyle\frac{-\log(u)}{\lambda \exp(\omega_1 \alpha_i + \omega_2 \phi_{ik})} \right)^{\frac{1}{p}} & \text{if} \ -\log(u) < b\\
    \left( \displaystyle\frac{-\log(u) -\lambda \exp(\omega_1 \alpha_i + \omega_2 \phi_{ik}) t_J^{p} + \lambda \exp(\nu + \omega_1 \alpha_i + \omega_2 \phi_{ik}) t_J^{p}}{\lambda \exp(\nu + \omega_1 \alpha_i + \omega_2 \phi_{ik})} \right)^{\frac{1}{p}}  & \text{if} \ -\log(u) \ge b
  \end{cases}
\end{equation}
with, again, $u \sim \text{Unif}(0, 1)$.
Similar closed-form expression could be obtained for exponential or Gompertz baseline hazard functions, again, following the work of Austin. \cite{Austin_2012}
We provide an R package with our implementation of this algorithm, which is openly available online on GitHub at \url{https://github.com/RedDoorAnalytics/simswjm}.

\begin{algorithm}
  \caption{Steps of the algorithm used to simulate data from the joint model of Equations \eqref{eq:jm-ci}, \eqref{eq:jm-gi} \label{alg:simjm}}

  \textbf{Input:} Data-generating parameters and study design settings

  \textbf{Step 1:} Simulate longitudinal data from a given stepped wedge design

  \begin{algorithmic}[1]
    \State Define the stepped wedge design: number of subjects per cluster, time periods, and treatment assignments;
    \State Simulate the stepped wedge design by assembling the number of clusters, subjects per cluster, and treatment periods according to the previous step;
    \State Draw the cluster-level random effects $\alpha \sim N(0, \sigma^2_{\alpha})$, for every cluster;
    \State Draw the subject-level random effects $\phi \sim N(0, \sigma^2_{\phi})$, for every study subject;
    \State Draw the residual errors $\varepsilon \sim N(0, \sigma^2_{\varepsilon})$, for every observation of every study subject;
    \State Calculate $Y$ for each subject in the trial by plugging the fixed model parameters (e.g., those that we described in Section \ref{mcsim-dgm}) and the simulated random effects and residual errors into the longitudinal submodel of Equations \eqref{eq:jm-ci}, \eqref{eq:jm-gi}.
  \end{algorithmic}

  \textbf{Step 2:} Simulate time to dropout from the study

  \begin{algorithmic}[1]
    \State Draw $u \sim \text{Unif}(0, 1)$ for every study subject;
    \State Calculate $b = \lambda \exp(\omega_1 \alpha_i + \omega_2 \alpha_{ik}) {t_J}^{p}$ according to the assumed data-generating parameters (e.g., those that we described in Section \ref{mcsim-dgm});
    \State Simulate dropout times $T^s$ by plugging $u, b$ and the data-generating parameters into Equation \eqref{eq:sim-weibull-simple};
    \State Generate dropout event indicator variable by comparing the simulated dropout time $T^s$ with the maximum allowed follow-up time (e.g., the time of the last measurement).
  \end{algorithmic}

  \textbf{Step 3:} Combine longitudinal and dropout data and censor longitudinal outcomes accordingly

  \begin{algorithmic}[1]
    \State Combine simulated data from \textbf{step 1} and \textbf{step 2};
    \State Censor simulated $Y$ values that are beyond dropout times $T^s$;
    \State Create start-stop notation for the survival submodel, including event indicator variables, required to account for the time-varying exogenous treatment.
  \end{algorithmic}

  \textbf{Output:} A simulated dataset for a stepped wedge trial under the joint model of Equations \eqref{eq:jm-ci}, \eqref{eq:jm-gi}

\end{algorithm}

Finally, we extend one of the simulation scenarios described above to test a few more settings in a non-factorial fashion (given the extensive number of simulation scenarios already implemented).
Specifically, we start with a scenario with informative dropout ($\nu = -0.2, \omega_1 = \omega_2 = \log(0.9)$) and with a non-zero treatment effect ($\delta = 5.0$).
Then, we update it by (1) reducing the number of clusters randomized to each intervention sequence to 3 (instead of 8), (2) reducing to 3 clusters per intervention sequence but doubling the number of participants per cluster to 100, (3) reducing the ICCs values by halving the variance of the cluster- and subject-specific random intercepts ($\sigma^2_{\alpha} = 1, \sigma^2_{\phi} = 27.5, \rho_{\alpha} = 0.416, \rho_d = 0.015$), and (4) increasing the ICCs values by doubling the random intercept variances ($\sigma^2_{\alpha} = 4, \sigma^2_{\phi} = 110, \rho_{\alpha} = 0.740, \rho_d = 0.026$), for a total of four additional scenarios.

We also extend the same scenario with informative dropout by testing an alternative, neutral dropout mechanism based on a mixed-effects logistic regression model (instead of time to dropout as described in step 2 of Algorithm \ref{alg:simjm}); note that, in this scenario, both models are misspecified.
Specifically, we assume that the probability of dropping out during the j\ith discrete period for the k\ith participants in the i\ith cluster is determined by the following model:
$$
P_{ijk}(\text{dropout}) = \text{logit}^{-1}(\gamma + \nu X_i + \omega_1 \alpha_i + \omega_2 \phi_{ik}),
$$
where $\gamma = -1.5$ is an intercept term and $\nu, \omega_1, \omega_2$ are parameters equivalent to those of the joint model (that we assume taking the same values described above).
The linear predictor (net of the intercept term) is standardized before simulating from this model, and we provide an easy-to-use implementation for simulating data under this dropout mechanism as well at \url{https://github.com/RedDoorAnalytics/simswjm}.

\subsection{Estimands} \label{mcsim-estimands}

The estimands of primary interest for this simulation study are the treatment effect parameters on the longitudinal outcome, denoted with $\delta$ (for simulation study \emph{A}) and with $\delta_0, \delta_1, \delta_2, \delta_3$ (for simulation study \emph{B}), and the period effects (the $\beta_j$ coefficients, $\forall$ values of $j$).
Of secondary interest, we also consider the within-individual and between-individuals ICCs $\rho_a$, $\rho_d$.
True values of each estimand were introduced in Section \ref{mcsim-dgm}.

\subsection{Methods} \label{mcsim-methods}

We evaluated the performance of the mixed-effects model for stepped wedge trials (introduced in Section \ref{mixed}) not taking into account informative dropout versus the shared random effects joint model introduced in Section \ref{joint-model}, assuming either a constant intervention model or a general time on treatment model for the longitudinal outcome.
Thus, we fit and compare the linear mixed models formalized in Equations \eqref{eq:sm-ne-ci} and \eqref{eq:sm-ne-gi} (simulation study \emph{A}), and the joint models formalized in Equations \eqref{eq:jm-ci} and \eqref{eq:jm-gi} (simulation study \emph{B}).

\subsection{Performance measures} \label{mcsim-performance}

The key performance measure of interest was bias, which quantifies whether a method targets the true value of a parameter on average.
We report bias on both an absolute and relative scale, the latter only for estimands with a true value that is not zero.
Additionally, we report empirical and model-based standard errors, to assess the performance of variance estimators and to compare the efficiency of the two methods, and convergence rates and coverage probabilities, for completeness.
All performance measures are described in more detail by Morris et al. \cite{Morris_2019}
Finally, note that model fits that did not converge are discarded when computing the performance measures of interest.

\subsection{Number of repetitions} \label{mcsim-nsim}

To determine how many repetitions were needed to estimate the performance measures of interest with sufficient precision, we first ran an arbitrary amount of 50 iterations.
We then analyzed these preliminary results and concluded that, with 1000 repetitions, we would expect a largest Monte Carlo error for bias of about 2\% of the absolute treatment effect value (either of $\delta$ or $\delta_0, \delta_1, \delta_2, \delta_3$), which we consider reasonable.
This procedure is described in more detail in the supplementary material.

\subsection{Software} \label{mcsim-software}

The simulation algorithms were implemented in R, while the analysis code for both models (linear mixed and joint model) was implemented in Stata using the built-in \texttt{gsem} command. \cite{R, Stata}
The simulation study is summarized in R using the \texttt{rsimsum} package. \cite{Gasparini2018}
All statistical code to replicate our simulations or fit the joint model using Stata is publicly available on GitHub at \url{https://github.com/ellessenne/swjm}, and is archived on Zenodo as well. \cite{paper_code}

\subsection{Results} \label{mcsim-results}

We report the results of the simulation study for the constant intervention model in \ref{mcsim-results-ci}, and the results for the general time on treatment model in Section \ref{mcsim-results-gi}.

\subsubsection{Results: simulation study \emph{A}, constant intervention model} \label{mcsim-results-ci}

First, we report on the convergence rates of the fitting procedures.
Overall, the joint model had lower convergence rates than the mixed model, which was expected given the extra computational complexity; nonetheless, convergence rates were acceptable in all scenarios with non-extreme dropout mechanisms, and we expect such extreme dropout mechanisms (with hazard ratios of 0.5 and 2.0) to be unlikely in practice.
These results are described in more detail in Appendix B of the supplementary material available online, and we discuss possible steps to improve convergence rates in practice in Appendix C. According to our experience, we expect these steps to lead to convergence of a joint model in most settings.

\begin{figure}
  \centering
  \includegraphics[width = \textwidth]{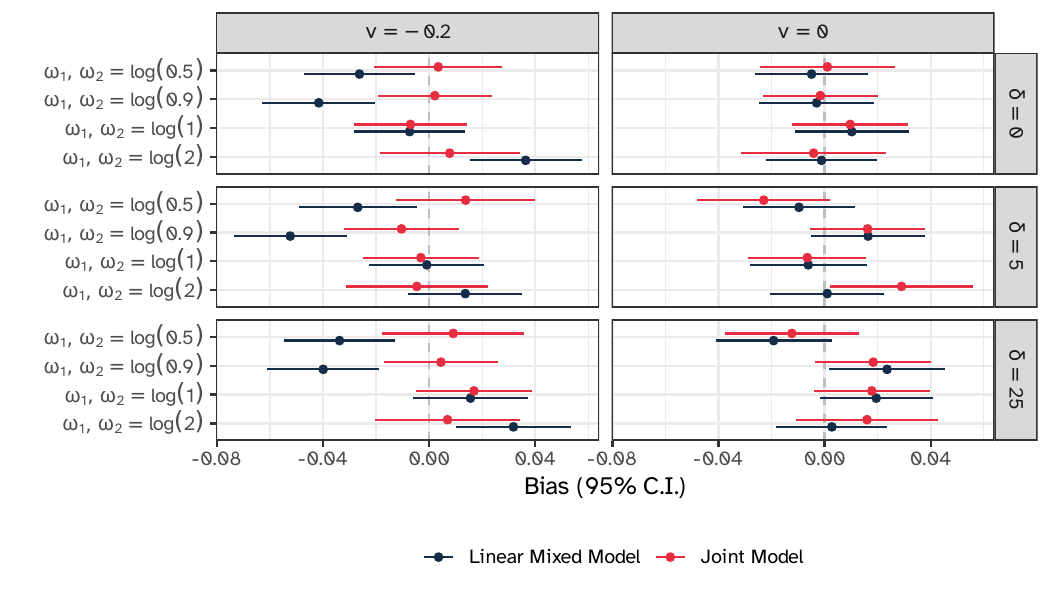}
  \caption{Estimated bias for the treatment effect $\delta$ on the longitudinal outcome with 95\% confidence intervals based on Monte Carlo standard errors, constant intervention model. \label{fig:p_bias_ci_delta}}
\end{figure}

Figure \ref{fig:p_bias_ci_delta} depicts the estimated bias for the treatment effect $\delta$ on the longitudinal outcome over repetitions and across all simulation scenarios and models; results are also tabulated in Appendix B of the supplementary material.
Overall, the joint model outperformed the linear mixed model, with negligible bias (if any) under every scenario.
Conversely, the linear mixed-effects model showed bias estimates in the scenarios with $\nu = -0.2$ and informative dropout, irrespective of the true treatment effect value.
These biases were, however, of small magnitude, as relative bias did not exceed 1\% (in absolute terms; Appendix B).
In all scenarios with $\nu = 0.0$ both models yielded unbiased results, and, as expected, both models performed well in all scenarios with non-informative dropout ($\omega_1 = \omega_2 = \log(1.0)$).

Interestingly, coverage probability was close to nominal across all scenarios and for both models, possibly because the relative bias to the true treatment effect under the mixed model is not substantial enough to bias statistical inference.
These results are reported in Appendix B of the supplementary material.

For the period effects, the linear mixed model showed significant bias at all possible time points when dropout was informative (Appendix B, supplementary material), up to 20\% (on a relative scale); the joint model outperformed the linear mixed model, with slight bias (<5\%) only in the extreme scenarios with $\omega_1 = \omega_2 = \log(0.5), \log(2.0)$.
Accordingly, coverage probability was poor when period coefficients were biased for both the linear mixed model and the joint model, and optimal otherwise.
As with the treatment effect $\delta$, when dropout was not informative, both models could estimate unbiased period effects with optimal coverage probabilities.

\begin{figure}
  \centering
  \includegraphics[width = \textwidth]{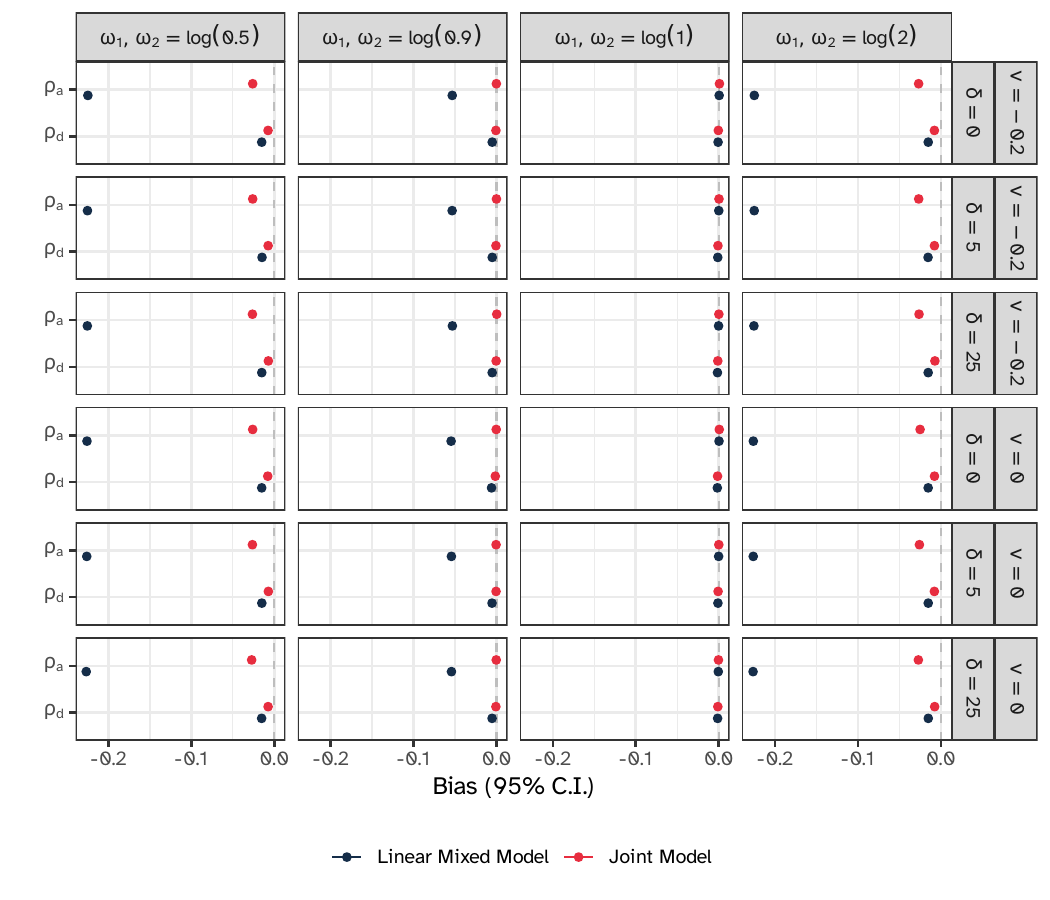}
  \caption{Estimated bias for the ICC parameters with 95\% confidence intervals based on Monte Carlo standard errors, constant intervention model. \label{fig:p_bias_ci_icc}}
\end{figure}

For the ICC, the joint model outperformed the linear mixed model in all scenarios with informative dropout, showing no bias or small bias (at worst, in some extreme scenarios), as in Figure \ref{fig:p_bias_ci_icc}; accordingly, coverage was low (Appendix B, supplementary material).
The under-coverage is not unexpected because typically, nominal coverage for such second-order parameters may require a large number of clusters, as demonstrated in previous simulations \cite{li_marginal_2022}.
More results on the estimation of the specific variance components are included, once again, in Appendix B of the supplementary material.

\subsubsection{Results: simulation study \emph{B}, general time on treatment model \label{mcsim-results-gi}}

\begin{figure}
  \centering
  \includegraphics[width = \textwidth]{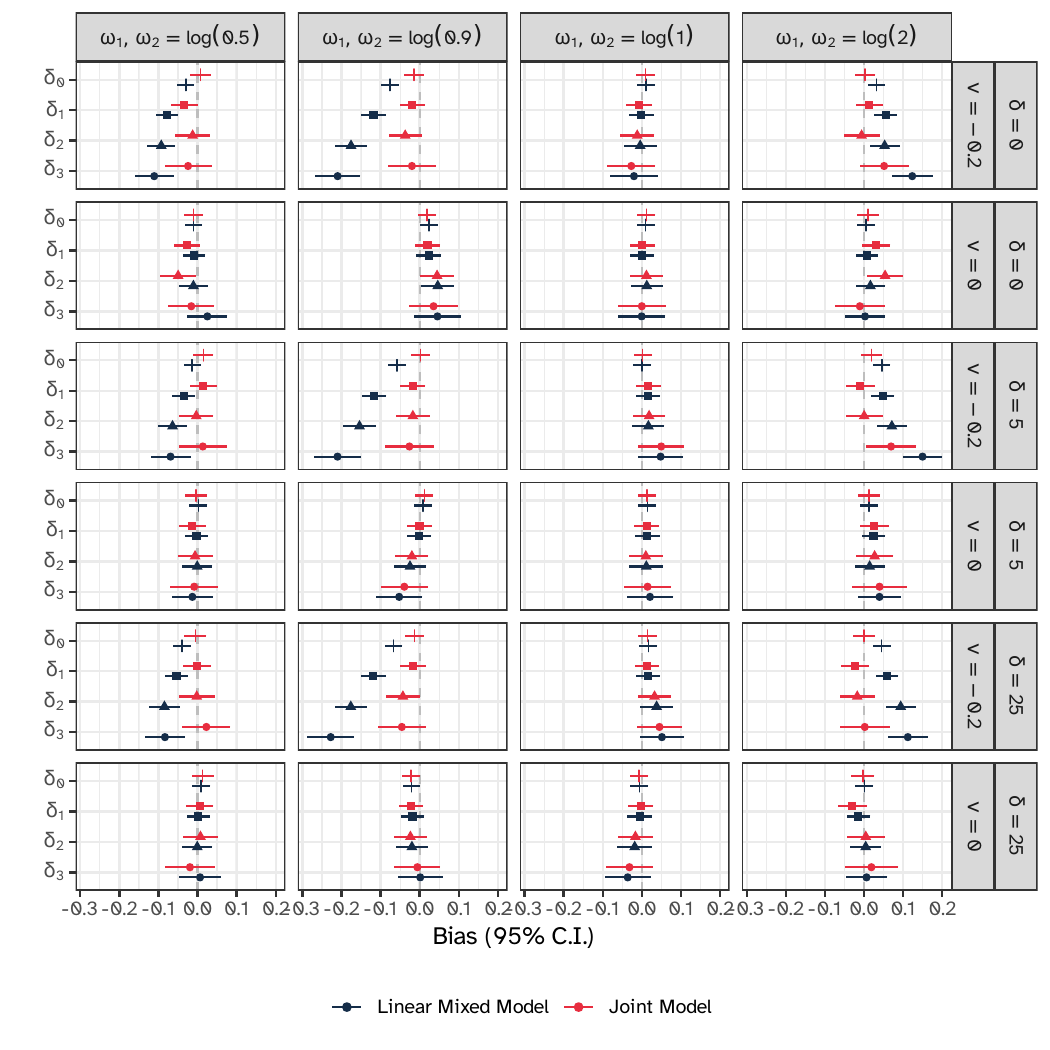}
  \caption{Estimated bias for the treatment effects $\delta_0, \delta_1, \delta_2, \delta_3$ on the longitudinal outcome with 95\% confidence intervals based on Monte Carlo standard errors, general time on treatment model. \label{fig:p_bias_gi_delta}}
\end{figure}

Figure \ref{fig:p_bias_gi_delta} depicts the estimated bias for the treatment effects $\delta_0, \delta_1, \delta_2, \delta_3$ on the longitudinal outcome over repetitions and across all simulation scenarios and models.
These results are tabulated in Appendix B of the supplementary material.
Overall, the joint model outperformed the linear mixed model, with no apparent bias across the board; biases in larger magnitude were observed for the treatment effects under higher level of exposure time; this is as expected because the effective sample size for estimating those parameters is smaller and the impact of informative data thinning due to dropout is more pronounced.
Scenarios that were not informative ($\omega_1 = \omega_2 = \log(1.0)$) showed no bias for either model; moreover, the mixed model performed well in scenarios with $\nu = 0.0$ as well, across values of the association parameters.
Coverage probability was close to nominal for both models.

Bias for the period effects and ICCs followed the same patterns described in Section \ref{mcsim-results-ci}. That is, (1) the joint model outperformed the linear mixed model across the board when dropout was informative and (2) both models returned unbiased estimates when dropout was not informative.

Overall, compared to the results summarized in Section \ref{mcsim-results-ci}, the benefits of the joint modeling approach are more evident under the general time on treatment model.
More results for the simulation under the general time on treatment model are included in Appendix B of the supplementary material available online.

\subsubsection{Results: additional simulation scenarios for both model formulations} \label{mcsim-results-extra}

\begin{figure}
  \centering
  \includegraphics[width = \textwidth]{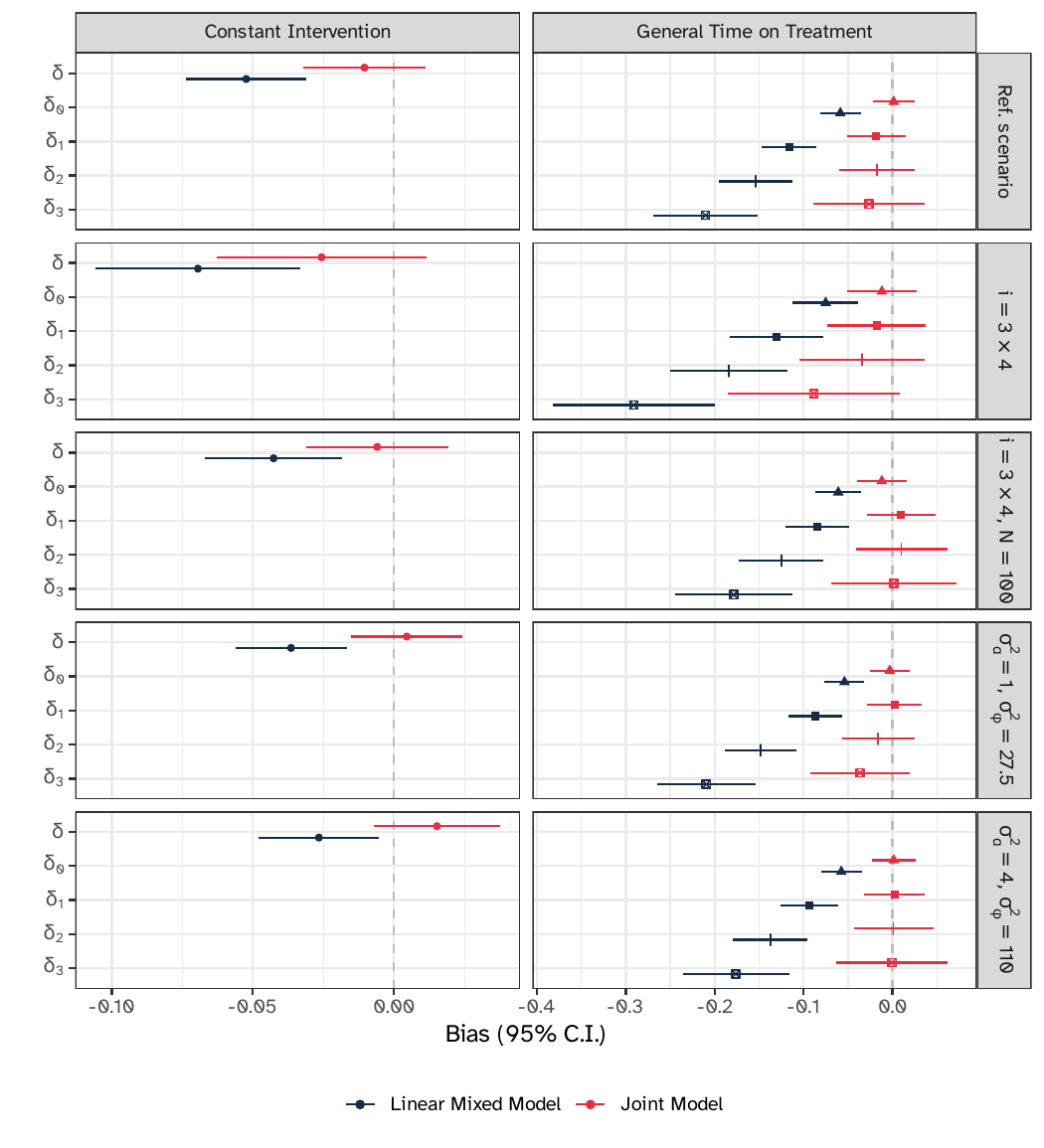}
  \caption{Estimated bias for the treatment effect $\delta$ on the longitudinal outcome with 95\% confidence intervals based on Monte Carlo standard errors, additional simulation scenarios for the constant intervention and general time on treatment models. Note that $i$ denotes the total number of clusters, and $N$ denotes the number of participants per cluster. \label{fig:p_bias_cigi_delta_extra}}
\end{figure}

Estimated bias for the treatment effect parameter(s) and the additional (non neutral) simulation scenarios are depicted in Figure \ref{fig:p_bias_cigi_delta_extra}.
The linear mixed model yielded biased estimates under every additional scenario, with greater bias in the scenario with fewer clusters per intervention sequence.
Superiority of the joint model was confirmed under every one of these additional scenarios with informative dropout: additional results are included in Appendix B of the supplementary material available online.

Finally, both models yielded biased estimates in the neutral simulation scenario: this is likely because both models are misspecified by the simulation design in such setting.
Nonetheless, the joint model still outperformed the linear mixed model across the board, with less bias for all model parameters (including ICCs and variance components).
These results are summarized in Appendix B of the supplementary material.

%%% ---------------

\section{Reanalysis of the `Frail Older Adults: Care in Transition' trial} \label{act}

The `Frail Older Adults: Care in Transition' (ACT) trial is a 24-month stepped wedge cluster randomized controlled trial that was conducted between May 2010 and March 2013 in 35 primary care practices in the Netherlands \cite{Hoogendijk_2016}.
The study included 1147 frail older adults, and aimed to study the impact of the Geriatric Care Model (GCM) on their quality of life; the study did not, however, show statistically significant beneficial effects of the GCM.
One potential issue with the ACT study is dropout from the study: the study participants were frail older adults with an average age of 80 years at baseline, which are at high risk of dropping out of the study, e.g., because of death.
In fact, after 24 months, more than 30\% of study participants had dropped out.

Thus, in this Section, we re-analyze data from the ACT trial using the joint modeling approach introduced in Section \ref{joint-model} to account for potentially informative dropout from the study; we also include results from the standard linear mixed model analysis for comparison purposes.
Specifically, we focus on two longitudinal outcomes: the mental health component score (MCS) and physical health component score (PCS) of the 12-item Short Form questionnaire (SF-12), which quantify quality of life.
We fit joint models assuming a constant intervention effect or a general time on treatment effect, equivalent to those introduced in Equations \eqref{eq:jm-ci} and \eqref{eq:jm-gi}, respectively, and adjust the longitudinal submodels for age, sex, and baseline variables on which the allocation groups differed at baseline (educational level, region and frailty index score), as in Hoogendijk et al. \cite{Hoogendijk_2016}
The survival submodel included time-varying treatment as a covariate and assumed the shared random effects association structure.
Comparable linear mixed models were fit by ignoring the time-to-event component.

\begin{figure}
  \centering
  \includegraphics[width = \textwidth]{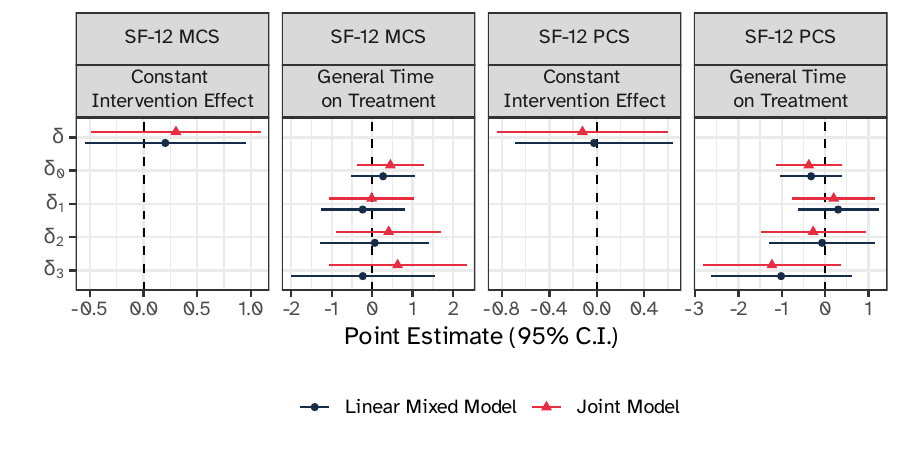}
  \caption{Estimated treatment effect parameters for the GCM on SF-12 MCS and PCS outcomes, ACT trial, according to both the constant intervention and general time on treatment joint and linear mixed models, with 95\% confidence intervals. \label{fig:act_delta}}
\end{figure}

The estimated treatment effect parameters are depicted in Figure \ref{fig:act_delta}: despite explicitly modeling the dropout process, we still estimated a small but not clinically meaningful association between the GCM and quality of life as measured by SF-12.
This estimated association was not statistically significant.
Overall, we observed treatment effect estimates closer to the null under the linear mixed model as compared to the joint model, but differences between the two were minor and did not change the interpretation of the results.
Estimated period effects, variance components, and, therefore, ICC values were also close when comparing the joint and linear mixed models, as illustrated in Appendix D of the Supplementary Material available online.

Due to the minor differences between estimates from the two models, we infer that dropout from the study in ACT is likely not strongly informative; we can test this assumption by studying the estimated coefficients of the survival submodel, which are listed in Table \ref{tab:act_tsurv}.
Specifically, the estimated parameter $\nu$ suggests a lower dropout rate for treated individuals, highlighting a direct effect of treatment on the hazard of dropout.
Moreover, the estimated $\omega_1$ parameters are close to zero and with large estimated standard errors, thus likely suggesting no association between the cluster-specific random intercept and dropout. In contrast, the estimated $\omega_2$ parameter suggests a significant association between the subject-specific random intercept and dropout but of small magnitude (relative to the size of the estimated variances; Appendix D, Supplementary Material) for the SF-12 MCS outcome, where the estimated $\omega_2$ parameter is approximately zero for the SF-12 PCS outcome.
Overall, the estimated values of $\omega_1, \omega_2$ suggest that either there is no differential dropout, e.g., between cluster or study subjects, or that this effect is of small magnitude leading to only minor impact on the final treatment effect estimates on the SF-12 outcomes.

In this example, the use of the joint modeling provides a useful sensitivity analysis for the ACT trial regarding the issue of possibly informative dropout.
In conclusion, our analyses suggest that the estimated treatment effects from the mixed model are only slight weaker but generally close to those from the joint model, likely because the dropout is not strongly informative.

\begin{table}
\centering
\caption{Estimated parameters of survival sub-model for the constant intervention and general time on treatment joint model, with 95\% confidence intervals. Note that parameters $\nu$, $\omega_1$, and $\omega_2$ can be interpreted as log hazard ratios. \label{tab:act_tsurv}}
\centering
\resizebox{\ifdim\width>\linewidth\linewidth\else\width\fi}{!}{
\begin{tabular}[t]{ccc}
\toprule
Parameter & Constant Treatment Effect & General Time on Treatment\\
\midrule
\addlinespace[0.3em]
\multicolumn{3}{l}{\textit{Outcome: SF-12 MCS}}\\
\hspace{1em}$\nu$ & -0.415 (-0.682, -0.149) & -0.416 (-0.682, -0.149)\\
\hspace{1em}$\omega_1$ & -0.305 (-0.831, 0.221) & -0.303 (-0.822, 0.216)\\
\hspace{1em}$\omega_2$ & -0.035 (-0.067, -0.002) & -0.035 (-0.067, -0.002)\\
\hspace{1em}$\log(\lambda)$ & -1.475 (-1.607, -1.344) & -1.475 (-1.607, -1.343)\\
\hspace{1em}$\log(p)$ & -0.065 (-0.174, 0.044) & -0.064 (-0.173, 0.044)\\
\addlinespace[0.3em]
\multicolumn{3}{l}{\textit{Outcome: SF-12 PCS}}\\
\hspace{1em}$\nu$ & -0.382 (-0.641, -0.123) & -0.387 (-0.647, -0.126)\\
\hspace{1em}$\omega_1$ & 0.061 (-0.074, 0.197) & 0.066 (-0.067, 0.200)\\
\hspace{1em}$\omega_2$ & -0.000 (-0.040, 0.040) & -0.001 (-0.041, 0.040)\\
\hspace{1em}$\log(\lambda)$ & -1.452 (-1.580, -1.324) & -1.451 (-1.579, -1.323)\\
\hspace{1em}$\log(p)$ & -0.076 (-0.184, 0.032) & -0.075 (-0.183, 0.033)\\
\bottomrule
\end{tabular}}
\end{table}

%%% ---------------

\section{Extensions of the joint model to other outcome types and association structures} \label{extensions}

The joint modeling framework can be extended in a variety of ways.
In this Section, we briefly introduce and discuss two of these: the extension to generalized outcome types (such as binary or count) and the extension to additional, more interpretable association structures between the longitudinal and time-to-event submodels.

Starting with the extension to generalized outcome types, we can extend the linear mixed model of Equation \eqref{eq:sm-ne-ci} (linear mixed model with a constant treatment effect) to a generalized linear mixed model for the mean of a certain outcome $Y_{ijk}(t)$:
\begin{equation}
  \label{eq:glmm-ci}
  g[E(Y_{ijk}(t))] = m_{ijk}(t) = \sum_{j=1}^J \beta_j I(T_{j - 1} < t \le T_j) + \delta X_i(t) + \alpha_i + \phi_{ik}
\end{equation}

In Equation \eqref{eq:glmm-ci}, $g(\cdot)$ denotes a certain link function; for instance, if we assume the identity function as a link function, we obtain the usual linear mixed model (as in Equation \eqref{eq:sm-ne-ci}).
Other possible link functions are the logit ($g(\cdot) = \logit(\cdot)$), probit ($g(\cdot) = \Phi(\cdot)$, with $\Phi$ denoting the cumulative distribution function of a standard normal distribution), and log ($g(\cdot) = \log(\cdot)$) functions, among others.

By framing the outcome within a generalized linear mixed modeling framework, the response (outcome) variable is no longer constrained to be continuous and can now follow any distribution belonging to the exponential family, such as binomial, negative binomial, Poisson, beta, or gamma.
For instance, for a stepped wedge trial with a binary outcome, we can thus assume a logit link function (and a binomial outcome distribution) to estimate log odds and log odds ratios.
For a count outcome, we can assume a log link and a Poisson distribution to estimate log expected outcome counts.

The extension to the joint modeling settings follows, by replacing the longitudinal submodel with a generalized linear mixed model:
\begin{equation}
\label{eq:gjm-ci}
\begin{cases}
g(E[Y_{ijk}(t)]) = \sum_{j=1}^J \beta_j I(T_{j - 1} < t \le T_j) + \delta X_i(t) + \alpha_i + \phi_{ik} \\
\lambda_{ijk}(t) = \lambda_0(t) \exp(\nu X_i(t) + \omega Z_{ik})
\end{cases}
\end{equation}

The newly-defined joint model of Equation \eqref{eq:gjm-ci} can now be used to model outcomes of any type in stepped wedge trials with (potentially) informative dropout.

Similar extensions can be defined for the general time on treatment models of Equations \eqref{eq:sm-ne-gi} and \eqref{eq:jm-gi}; these are omitted from this Section for simplicity.
Note that these generalized joint models can be fitted using the \texttt{gsem} command in Stata; \cite{Stata} every built-in distribution for the longitudinal outcome is supported within this framework, with a certain amount of additional computational complexity depending on the distribution of choice.
Example statistical code to fit these joint models is included on GitHub at \url{https://github.com/RedDoorAnalytics/simswjm}.

For the second proposed extension, recall that in every joint model introduced so far, we assumed the shared random effects parametrization for the association structure between the longitudinal and dropout submodels.
The interpretation of the estimated association parameters reflects the change in the log hazard of dropout for a unit change in the deviation of each cluster/subject from the population mean; note that this association structure is time-independent, as the latent random intercept values do not change over time.
Alternative, time-dependent association structures have been proposed in the literature. \cite{gould_joint_2015, papageorgiou_overview_2019}
One such alternative association structure is the so-called \emph{expected value association structure}, where we link the two submodels via the expected value of the longitudinal outcome at a certain point in time $t$.
For instance, the joint model of Equation \eqref{eq:jm-ci} could be re-written as:
\begin{equation}
  \label{eq:jm-ev-ci}
  \begin{cases}
  Y_{ijk}(t) = \sum_{j=1}^J \beta_j I(T_{j - 1} < t \le T_j) + \delta X_i(t) + \alpha_i + \phi_{ik} + \varepsilon_{ijk}(t) \\
  \lambda_{ijk}(t) = \lambda_0(t) \exp(\nu X_i(t) + \omega E[Y_{ijk}(t)])
  \end{cases}
\end{equation}
where $E[Y_{ijk}(t)]$ denotes the expected value of the outcome $Y$ at time $t$.
The association parameter $\omega$ can now be interpreted as the change in log hazard of dropout for a unit change in the expected value of the outcome $Y$ at time $t$.
Note that the expected outcome value changes over time, which introduces additional computational complexity in the estimation process.
Alternative association structures are discussed elsewhere and include, among others, the rate of change in the longitudinal outcome at time $t$ or the (possibly weighted) cumulative effect. \cite{papageorgiou_overview_2019}
Alternative association structures are not yet available within Stata's \texttt{gsem} command and would require ad-hoc software development to accommodate all the additional complexities of stepped wedge trials (such as the hierarchical structure and the time-varying exogenous treatment assignment) before being usable in practice.

%%% ---------------

\section{Discussion} \label{discussion}

%\textcolor{red}{\emph{Part 1: Summary of the manuscript:}}

We introduced a joint longitudinal-survival model that can be used in the settings of closed-cohort stepped wedge CRTs to account for non-ignorable dropout.
The proposed model can be used with both constant and general time on treatment effect parametrizations, can accommodate time-varying exogenous covariates in the dropout submodel (such as treatment), and can accommodate different ways of linking the submodels (e.g., the user-defined form of $\omega$ in Equation \eqref{eq:jm-ci}).
Moreover, we provide annotated technical guidance and statistical code for readers to apply the methodology in practice, including synthetic datasets --- this is openly available on GitHub at \url{https://github.com/ellessenne/swjm} and archived on Zenodo as well. \cite{paper_code}
Usual model-building strategies apply, e.g., in terms of functional form of covariates and interaction terms, baseline hazard distributions, random effects specification.
Fit of the joint model with the observed data can be assessed by comparing different model formulations using, e.g., information criteria (such as the Akaike and Bayesian Information Criteria \cite{akaike_information_1973, schwarz_estimating_1978, stoica_model_order_2004}) or by comparing predicted and observed values.
Robustness of the random effects assumptions increases with the number of observations per cluster; \cite{gould_joint_2015} in small sample size settings, $t$-distributed random effects could be used instead, as long as the software implementation supports it. \cite{crowther_extended_2017}

Moreover, we developed an algorithm to simulate data from stepped wedge CRTs with or without informative dropout, which can be useful beyond the settings of this manuscript, e.g., to estimate the required sample size for a future study using a simulation-based approach.
The code is openly available for other to re-use and apply at \url{https://github.com/RedDoorAnalytics/simswjm}.

Using Monte Carlo simulation and realistic data-generating mechanisms, we showed that the joint model outperformed the standard linear mixed-effects model in a variety of scenarios with informative dropout.
Specifically, the joint model yielded unbiased treatment and period effect estimates, with the mixed model performing significantly worse in estimating period effects and treatment effects under the general time on treatment parametrization.
Moreover, the mixed model yielded significant bias in the estimated variance components and, therefore, the ICCs.
The mixed model was, however, relatively robust in some of the scenarios with informative dropout, with unbiased treatment effect estimates at the cost of bias in the estimated variance components, which were inflated to accommodate the additional heterogeneity due to informative dropout.
Interestingly, fitting the joint model in scenarios where dropout was not informative did not introduce any bias in the analysis.
This highlights that fitting the joint model is a viable strategy to confirm the results of a mixed model when it is unclear whether dropout from the study is informative or not, as a possible sensitivity analysis.

This manuscript has some strengths and limitations.
Among the strengths, our Monte Carlo simulation study is structured to follow the ADEMP framework and used realistic data-generating mechanisms informed by the ACT study, which was re-analyzed in Section \ref{act}.
The joint modeling approach can be implemented using off-the-shelf statistical software and can be easily extended to accommodate other variants of the shared random effects association structure, different treatment effects parametrizations (e.g., the linear time-on-treatment and delayed treatment effect parametrizations discussed by Li et al. \cite{Li_2021}), and baseline hazard functions for the dropout submodel.
These extensions can be easily implemented using \texttt{gsem} in Stata.
Our approach can accommodate time-varying, exogenous covariates in the survival submodel, which is not trivial; as we demonstrate how to fit this with off-the-shelf statistical software, this can be easily extended to other settings, study designs, and disease areas.
We also developed an algorithm for simulating data from the joint model in the settings of stepped wedge CRTs; we implemented this in an R package that is available on GitHub at \url{https://github.com/RedDoorAnalytics/simswjm}, with statistical code to replicate our simulation study and implement the joint model in practice also available on GitHub at \url{https://github.com/ellessenne/swjm}.

Among the limitations of our study, the main data-generating mechanisms are based on the joint model, therefore the good performance of the joint model is somewhat expected. \cite{niesl_overoptimism_2022}
Nonetheless, superiority of the joint model was confirmed in a neutral simulation scenario that we tested, and the results of this manuscript are in agreement with the current literature on joint longitudinal-survival modeling, while providing new evidence in the settings of hierarchical models with more that one level of nesting. \cite{cuer_handling_2020, cuer_joint_2021}
Thomadakis et al. showed that comparable joint longitudinal-survival models could be biased in certain MAR scenarios, but we did not observe such behavior --- likely due to the non-informative scenarios reducing to MCAR, where joint models are expected to estimate unbiased results. \cite{thomadakis_longitudinal_2019, njagi_characterization_2014}
The results of the simulation study can only be generalized to settings close to the assumed data-generating mechanisms.
Implementing and reporting a larger number of scenarios would likely become unwieldy, and further simulation studies are therefore warranted.
Next, we modeled the dropout process in continuous time (compared to, e.g., discrete-time models as in Wang and Chinchilli \cite{wang_analysis_2022}), therefore we need to assume that dropout times are known nearly exactly.
Of note, our methods can be directly applied to the individually-randomized stepped wedge design, \cite{hooper2019improving} which is a recent variation of the traditional individually-randomized parallel-arm design by including staggered treatment assignment for efficiency improvement.
In that case, the appropriate joint model is a special case of the model of Equation \eqref{eq:jm-ci} after removing the cluster-level random effect $\alpha_i$, and we provide sample code to implement this procedure in the GitHub repository.
Finally, some of the extensions discussed throughout the manuscript and in Section \ref{extensions} require further software developments before they can be used in practice.
This is also left for future work.

This work can also be further extended beyond what was introduced and discussed in Section \ref{extensions}.
For instance, we currently use a frequentist approach with maximum likelihood estimation: a natural extension is to frame the joint model within a Bayesian framework, to allow the incorporation of prior information (e.g., from previous trials).
Moreover, we do not consider that some of the outcome measures are non-directly measurable, i.e., with studies using a set of observed indicators from questionnaires or measurement scales to evaluate an unobservable latent variable.
Saulnier et al. propose a latent process joint model to accommodate such outcomes, which could be combined with the methodology of this manuscript for the settings of stepped wedge trials. \cite{saulnier_joint_2022}
Finally, there exist more complicated random-effects structures for modeling longitudinal outcomes in closed-cohort stepped wedge designs, such as those that incorporate a random cluster-by-period effect and further allow for discrete-time correlation decay. \cite{li_design_2020, Li_2021}
These extensions will inevitably introduce additional computational complexity under the joint modeling framework and will be important directions for future research.

In conclusion, we described joint modeling methodology that can be used in the settings of stepped wedge cluster randomized trials when dropout from the study is suspected to be informative, and we showed that the joint modeling approach reduced bias across the board compared to standard linear mixed-effects models that ignore the dropout process.
Annotated statistical code is provided for readers to apply this methodology in practice.

% Backmatter

\bmsection*{Author contributions}

A. G.: conceptualization, methodology, software, writing - original draft, writing - review \& editing, visualization, formal analysis, data curation, project administration.
F. L.: conceptualization, methodology, writing - review \& editing, supervision, funding acquisition.
E. O. H: resources (provision of ACT data), writing - review \& editing.
M. J. C.: methodology, software, writing - review \& editing, supervision.
M. O. H.: conceptualization, methodology, writing - review \& editing, supervision, funding acquisition.

\bmsection*{Acknowledgments}

The ACT study is funded by The Netherlands Organization for Health Research and Development (ZonMw): Dutch National Care for the Elderly Program grant number 311080201.
Research in this article was partially supported by the Patient-Centered Outcomes Research Institute\textsuperscript{\textregistered} (PCORI\textsuperscript{\textregistered} Awards ME-2020C1-19220 to M. O. H. and ME-2022C2-27676 to F. L.).
F. L. and M. O. H. are funded by the United States National Institutes of Health (NIH), National Heart, Lung, and Blood Institute (grant number R01-HL168202).
M.O.H. is also funded by grant number P30-HS029745, which is funded by the Agency for Healthcare Research and Quality (AHRQ) and PCORI\textsuperscript{\textregistered}.
All statements in this report, including its findings and conclusions, are solely those of the authors and do not necessarily represent the views of the NIH, AHRQ, or PCORI\textsuperscript{\textregistered} or its Board of Governors or Methodology Committee.

\bmsection*{Financial disclosure}

Nothing to disclose.

\bmsection*{Conflict of interest}

Nothing to disclose.

\bmsection*{Supporting information}

Additional results for the simulation studies reported in Section \ref{mcsim-results} and for the applied example reported in Section \ref{act} are included in the Supplementary Material, available online on the journal website.
Statistical code to replicate the simulation study and an R package implementing the simulation algorithms are available on GitHub at \url{https://github.com/ellessenne/swjm} and \url{https://github.com/RedDoorAnalytics/simswjm}, respectively.
Replication and supporting code is archived on Zenodo as well. \cite{paper_code}

\end{document}